\documentclass{sig-alternate.fix-ref-space}
\usepackage{xspace}
\usepackage{multirow}

\newcommand{\omt}[1]{}

\newcommand{\query}{q}
\newcommand{\doc}{d}
\newcommand{\offspringVar}{o}

\newcommand{\corpus}{{\cal C}}
\newcommand{\word}{w}
\newcommand{\wordseqlength}{n}
\newcommand{\wordseq}{\word_1 \word_2 \cdots \word_\wordseqlength}

\newcommand{\wordseqvar}{s}
\newcommand{\wordseqvarlength}{\vert \wordseqvar \vert}

\newcommand{\altArbGroupMember}{x}

\newcommand{\set}[1]{\{#1\}}
\newcommand{\definedas}{\stackrel{def}{=}}

\newcommand{\kld}[2]{D\left(#1 \; \Big\vert\Big\vert \,\, #2\right)}
\newtheorem{definition}{Definition}
\newcommand{\freq}[2]{{\rm tf}(#1 \in #2)}
\newcommand{\wordIndex}{j} 
\newcommand{\entropy}{H}

\newcommand{\relret}[1]{}

\newcommand{\prob}{p}

\newcommand{\ilmprob}{\prob} 
\newcommand{\baseprob}{\prob}

\newcommand{\genprob}[2]{\inducedprob{\ilmprob}{#1}{#2}}

\newcommand{\inducedprob}[3]{\ensuremath{#1_{#2}(#3)}}

\newcommand{\mlprob}[2]{\inducedprob{\ilmprob^{MLE}}{#1}{#2}}
\newcommand{\mlprobTerm}[2]{\inducedprob{\widetilde{\ilmprob}^{\,MLE}}{#1}{#2}}

\newcommand{\dirichletParam}{\mu}
\newcommand{\dirichletLM}[3]{\inducedprob{\ilmprob^{[\dirichletParam]}}{#1}{#2}}
\newcommand{\dirichletLMTerm}[3]{\inducedprob{\widetilde{\ilmprob}^{\;[\dirichletParam]}}{#1}{#2}}
\newcommand{\bareMLProb}[1]{\ilmprob^{MLE}_{#1}}

\newcommand{\bareDirProb}[1]{\ilmprob^{[\dirichletParam]}_{#1}}
\newcommand{\genprobKL}[2]{\prob^{KL,\dirichletParam}_{#1}(#2)}

\newcommand{\initTag}{{\rm init}}
\newcommand{\topRetGroup}{{\cal D}_{\initTag}}
\newcommand{\numretdocs}{\vert \topRetGroup \vert}

\newcommand{\score}{{\mathit Cen}}

\newcommand{\scoreFnG}[2]{\score_I(#1;#2)}
\newcommand{\scoreFnGR}[2]{{\it Cen}_{RI}(#1;#2)}

\newcommand{\scoreGeneric}{\score}

\newcommand{\propogatePSingleParam}[1]{\baseprob(\cdot)}

\newcommand{\prestigeText}{centrality\xspace}
\newcommand{\PrestigeText}{Centrality\xspace}
\newcommand{\prestigious}{central\xspace}

\newcommand{\pointTo}{\to}

\newcommand{\dampFactor}{\lambda} 

\newcommand{\arbGraph}{G}
\newcommand{\dgraphU}{G_U}
\newcommand{\dgraphW}{G_W}

\newcommand{\edgeAssert}[2]{#1 \pointTo #2}
\newcommand{\edge}[2]{(\edgeAssert{#1}{#2})}
\newcommand{\wtName}{{\mathit wt}}
\newcommand{\edgeWeight}[2]{\wtName\edge{#1}{#2}}
\newcommand{\edgeWeightDec}[3]{\wtName_{#3}\edge{#1}{#2}}
\newcommand{\smoothG}[2]{#1^{[#2]}}
\newcommand{\node}{v}
\newcommand{\nodeFrom}{\node_1}
\newcommand{\nodeTo}{\node_2}
\newcommand{\smoothedText}{smoothed\xspace}
\newcommand{\wtedInDegree}{influx\xspace}
\newcommand{\WtedInDegree}{Influx\xspace}

\newcommand{\numGenerators}{\alpha}

\newcommand{\generating}{generating\xspace}
\newcommand{\generation}{generation\xspace}
\newcommand{\generationText}{\generation}
\newcommand{\generationCaps}{Generation\xspace}

\newcommand{\generate}{generate\xspace}

\newcommand{\generated}{generated\xspace}
\newcommand{\generators}{{\mathit TopGen}}
\newcommand{\generatorsText}{generators\xspace}

\newcommand{\generatorText}{generator\xspace}
\newcommand{\docmember}{\doc}

\newcommand{\genGroupOfDoc}[1]{\generators(#1)}

\newcommand{\offText}{offspring\xspace}
\newcommand{\offsText}{\offText}
\newcommand{\offDoc}{\offspringVar}
\newcommand{\genDoc}{g}

\newcommand{\Scaled}{Recursive\xspace}

\newcommand{\firstmention}[1]{{\bf #1}}

\newcommand{\simpleCount}{{\em Uniform \WtedInDegree}\xspace}
\newcommand{\weightedCount}{{\em Weighted \WtedInDegree}\xspace}
\newcommand{\onlyUniformPrior}{{\em \Scaled Uniform \WtedInDegree}\xspace}
\newcommand{\onlyPrior}{{\em \Scaled Weighted \WtedInDegree}\xspace}
\newcommand{\asPrior}[1]{{#1}+{\em LM}\xspace}
\newcommand{\simpleCountPrior}{\asPrior{\simpleCount}}
\newcommand{\weightedCountPrior}{\asPrior{\weightedCount}}
\newcommand{\uniformRandomWalk}{\asPrior{\onlyUniformPrior}}
\newcommand{\weightedRandomWalk}{\asPrior{\onlyPrior}}

\newcommand{\inputAbbrev}{In}

\newcommand{\scaledAbbrev}{R}
\newcommand{\abbrevAsPrior}[1]{{#1}+LM}
\newcommand{\abbrevVote}{U-\inputAbbrev}
\newcommand{\abbrevWeightedVote}{W-\inputAbbrev}
\newcommand{\abbrevOnlyUniformPrior}{\scaledAbbrev-U-\inputAbbrev}
\newcommand{\abbrevOnlyPrior}{\scaledAbbrev-W-\inputAbbrev}
\newcommand{\abbrevUniformRandomWalk}{\abbrevAsPrior{\scaledAbbrev-U-\inputAbbrev}}
\newcommand{\abbrevWeightedRandomWalk}{\abbrevAsPrior{\scaledAbbrev-W-\inputAbbrev}}
\newcommand{\abbrevSimpleVotePrior}{\abbrevAsPrior{U-\inputAbbrev}}
\newcommand{\abbrevWeightedVotePrior}{\abbrevAsPrior{W-\inputAbbrev}}

\newcommand{\abbrevInit}{init. ranking}
\newcommand{\abbrevUniformPrior}{uniform (= init)}

\newcommand{\rerankApproach}{structural \mbox{re-rank}\-ing\xspace}
\newcommand{\RerankApproach}{Structural \mbox{re-rank}\-ing\xspace}

\newcommand{\ReRankApproachU}{Structural \mbox{Re-Rank}\-ing\xspace}

\newcommand{\tie}{~}
\newcommand{\usBetter}{\blacklozenge}
\newcommand{\themBetter}{{\rm \square}}
\newcommand{\A}{$\usBetter$} 
\newcommand{\Am}{$\usBetter$} 
\newcommand{\Amm}{$\usBetter$} 
\newcommand{\B}{$\tie$} 
\newcommand{\C}{$\tie$}
 
\newcommand{\Dm}{$\themBetter$} 
\newcommand{\F}{$\themBetter$} 
\newcommand{\X}{$\themBetter$}

\numberofauthors{2}

\title{PageRank without Hyperlinks: \ReRankApproachU using Links
  Induced by Language Models}

\conferenceinfo{SIGIR'05,}{August 15--19, 2005, Salvador, Brazil.}
\CopyrightYear{2005}
\crdata{1-59593-034-5/05/0008}

\begin{document}

\author{
\alignauthor Oren Kurland $^{1,3}$ \\
\email{kurland@cs.cornell.edu} \\[5mm]
\begin{tabular}{l}
\affaddr{$\quad$1.~Computer Science Department,
Cornell University, Ithaca NY 14853, U.S.A.}  \\
\affaddr{$\quad$2.~Language Technologies Institute,
Carnegie Mellon University, Pittsburgh PA 15213, U.S.A.}\\
\affaddr{$\quad$3.~Computer Science Department, Carnegie
Mellon University, Pittsburgh PA 15213, U.S.A.} 
\end{tabular}\\
\alignauthor Lillian Lee $^{1,2,3}$ \\
\email{llee@cs.cornell.edu}
}

\maketitle
\begin{abstract}
Inspired by the PageRank and HITS (hubs and authorities) algorithms
for Web search,
we propose a {\em \rerankApproach} approach 
to ad hoc information retrieval:
we reorder the documents in an initially retrieved set
by exploiting 
asymmetric
relationships between 
them.
Specifically, we consider
{\em \generationText links}, 
which
indicate
that the language model induced from one document 
assigns high probability to the text of another;
in doing so, we take care to prevent bias against long documents. 
We study a number of 
re-ranking criteria 
based on measures of {\em \prestigeText} 
in the
{graphs} formed by \generationText links,
and show that integrating \prestigeText into standard
language-model-based  retrieval
is quite effective at improving precision at top ranks.
\end{abstract}

\vspace{1mm}
\noindent
{\bf Categories and Subject Descriptors:} H.3.3 {[Information Search
and Retrieval]}: {Retrieval models}

\vspace{1mm}
\noindent
{\bf General Terms:} Algorithms, Experimentation

\vspace{1mm}
\noindent
{\bf Keywords:} language modeling, PageRank, HITS, hubs, authorities, social
networks, high-accuracy retrieval, graph-based retrieval,   structural re-ranking

\section{Introduction} 
\label{sec:intro} 

Information retrieval systems capable of achieving high precision at
the top ranks of the returned results would be of obvious benefit to
human users, and could also aid pseudo-feedback approaches,
question-answering systems, and  other applications that  use IR
engines for pre-processing purposes
\cite{Ruthven+Lalmas:03a,Tao+Zhai:04a, Shah+Croft:04a}.
But crafting such systems remains a key research challenge.

The PageRank Web-search algorithm \cite{Brin+Page:98a} 
uses
 explicitly-indicated 
inter-document
relationships 
as an
additional source of information beyond textual content, computing
which documents are the most {\em \prestigious}.
Here, we consider 
adapting this idea to
corpora 
in which explicit links between documents
do not exist.

How should we form links in a
non-hypertext setting?  
While previous work in
summarization has applied PageRank to 
cosine-based links \cite{Erkan+Radev:04b},
we draw on research demonstrating the success
of using {\em language models} to improve IR performance in general
\cite{Ponte+Croft:98a,Croft+Lafferty:03a} and to model inter-document
relationships in particular \cite{Kurland+Lee:04a}.
Specifically, we employ {\em \generationText links}, which are
based on
the probability assigned by the language model induced from one document
to the term sequence comprising another.
\footnote{
While the term ``\generate'' is convenient, we do not think of  a ``\generatorText'' document 
or
 language model as
literally 
``creating''
others.
Other work further discusses this issue and proposes alternate terminology
(e.g., ``render'') \cite{Kurland+Lee+Domshlak:05a}.
}  Our use of such links
echoes
the standard
language-model-based
ranking principle, first introduced in  
\cite{Ponte+Croft:98a},  that a document is relevant to the extent that its corresponding language model assigns high probability to the
query.  
However, given that we are working with multiple documents rather than
a single query, we employ a technique that compensates for
length bias in estimating \generationText probabilities.

We note that the analogy between hyperlinks and \generationText links
is not perfect. 
In
particular, one can attribute much of the success of link-based Web-search algorithms to the fact that hyperlinks are
(often) {human-provided} certifications that two pages are truly
related \cite{Kleinberg:99a}.
In contrast,
automatically-induced \generationText links 
are surely a noisier source of
information.  To compensate, 
we advocate
an approach 
(used elsewhere as well \cite{Willett:85a,Hearst+Pedersen:96a,Kleinberg:99a,Leuski:01a,Tombros+Villa+Rijsbergen:02a,Liu+Croft:04a})
 that we term
{\em \rerankApproach}:
we use inter-document relationships to  compute an ordering not of the
entire corpus,
but of 
a (possibly unranked)
set of documents produced by 
an
initial retrieval method.
This set 
should provide
a reasonable ratio of relevant to non-relevant
documents, and thus 
form a good  foundation for our 
algorithms.
Note that our approach differs in spirit
from 
pseudo-feedback-based methods \cite{Ruthven+Lalmas:03a},
which define a model based on 
the initially retrieved documents 
expressly  in order to 
re-rank the entire corpus.
Indeed, since the quality of the initially retrieved results 
plays a major role in determining the effectiveness of
pseudo-feedback-based algorithms \cite{Tao+Zhai:04a},
our methods
can potentially 
serve to greatly enhance the input to them.

To compute \prestigeText values for a given \generationText graph, we propose a number of methods,
including variants of PageRank 
\cite{Brin+Page:98a} and HITS
(a.k.a. hubs and authorities)
\cite{Kleinberg:99a}.
Comparisons on various TREC datasets 
against numerous baselines (including use of cosine-based links and
re-ranking employing only document-specific characteristics)
show that 
language-model-based re-ranking using \prestigeText
as a form of ``document prior'' 
is indeed
successful at 
moving relevant documents in the initial retrieval results higher up
in the list.

\section{Structural Re-Ranking}
\label{sec:algorithms}

Throughout this section, we assume that the following
 have been
fixed: the corpus 
$\corpus$ (in which each document has been assigned a unique numerical
ID); the query $\query$; 
the  set $\topRetGroup
\subseteq \corpus$ of top documents
returned by some initial retrieval algorithm in response to $\query$
 (this is the set upon which re-ranking is performed);
and the value of an {\em ancestry} parameter $\numGenerators$ that
pertains to our graph construction process. 

For each document $\doc \in \corpus$, $\genprob{\doc}{\cdot}$ denotes
the
smoothed
unigram language model induced from $\doc$ (estimation
details appear in Section \ref{sec:lmspec}).  We use $\genDoc$ and
$\offDoc$ to distinguish between a document treated as a ``\generatorText''
and a document treated as ``\offText'', that is, something that is
\generated
(details below).

We use the notation $(V,\wtName)$ for weighted directed
graphs: $V$ is the set of vertices and $\wtName: V \times V
\rightarrow \set{y \in \Re : y \geq 0}$ is the  {\em edge-weight
function}.  Thus, there is a directed
edge between every ordered pair of vertices,  but $\wtName$ may assign zero
weight to some edges.  We write
$\edgeWeight{\nodeFrom}{\nodeTo}$ to denote the value of $\wtName$
on edge $(\nodeFrom,\nodeTo)$.

\subsection{\generationCaps Graphs}
\label{sec:graph}

Our use of language models to form links  can be
motivated by considering
the following two documents:\smallskip \\ 
\begin{tabular}{p{.2in}ll}
& $\doc_1$: & Toronto Sheffield Salvador \\
& $\doc_2$: & Salvador Salvador Salvador
\end{tabular} \smallskip \\ 
Knowing that $\doc_2$ is important (i.e., \prestigious or
relevant) would provide strong evidence that $\doc_1$ is at
least somewhat important. However, knowing that $\doc_1$ is
very important does {\em not} allow us to conclude that $\doc_2$ is, since the
importance of $\doc_1$ might stem from its first two terms.  Using
{\em language models} induced from  documents 
enables us to
capture this asymmetry in how \prestigeText is
propagated:
we 
allow a document $\doc$ to receive support for \prestigeText status from a 
document $\offDoc$ 
only to the extent that  
$\genprob{\doc}{\offDoc}$ is relatively large.
(If  $\offDoc$ is not in fact important, the support it provides may
not be significant.)
Note that ranking 
documents 
by $\genprob{\doc}{\query}$,
as first proposed by Ponte and Croft \cite{Ponte+Croft:98a}, can be
considered 
a variation of this principle.

We are thus led to the following definitions.
\begin{definition}
\label{defn:docEnv}
The {\em top $\numGenerators$ \generatorsText} of a  document $\docmember \in
\topRetGroup$, denoted $\genGroupOfDoc{\docmember}$, is the set  of
$\numGenerators$ documents $\genDoc \in \topRetGroup-
\set{\docmember} $
that yield the highest
$\genprob{\genDoc}{\doc}$, where ties are broken by document ID. (We
suppress $\numGenerators$ in our notation for 
clarity.)
\end{definition}
\begin{definition}
The {\em \offsText} of a document $\doc \in
\topRetGroup$ are those documents that $\doc$ is a top
\generatorText of, i.e., the set
$\set{\offDoc \in \topRetGroup : \doc
  \in \genGroupOfDoc{\offDoc}}$.
\end{definition}
Note that multiple documents can share \offsText, and that
it is possible for a document to have no \offsText.

We can encode top-\generationText  relationships
using either of two 
{\em \generationText
  graphs} 
$\dgraphU = (\topRetGroup,\wtName_U)$ and $\dgraphW = (\topRetGroup,\wtName_W)$, where
for $\offDoc, \genDoc \in \topRetGroup$,
\begin{eqnarray*}
\edgeWeightDec{\offDoc}{\genDoc}{U} & = & \begin{cases}1 &  \text{ if } \genDoc \in
  \genGroupOfDoc{\offDoc}, \\  0 & \text{ otherwise};
\end{cases} \\
\edgeWeightDec{\offDoc}{\genDoc}{W} & = &  \begin{cases}
  \genprob{\genDoc}{\offDoc} & \text{ if } \genDoc \in
  \genGroupOfDoc{\offDoc}, \\ 0 & \text{ otherwise}.\end{cases}
\end{eqnarray*}
Thus, in both graphs, positive-weight edges lead only from 
\offsText to their respective top $\numGenerators$ \generatorsText; but
$\dgraphU$ treats (edges to) the top \generatorsText of  $\offDoc$
{\em uniformly}, whereas $\dgraphW$ differentially {\em weights}
them by the
probability their induced language models assign to $\offDoc$.

Some of our algorithms require
``smoothed'' 
versions of these graphs, in which 
all edges (including 
self-loops) 
have non-zero weight,
to
work correctly.  To be specific, we employ PageRank's
\cite{Brin+Page:98a} smoothing technique.
\begin{definition}
\label{defn:smoothGraph}
 Given 
an edge-weighted directed graph $G=(\topRetGroup,
\wtName)$
and 
smoothing parameter $\dampFactor \in [0,1)$,
the {\em \smoothedText} graph 
$\smoothG{G}{\dampFactor}=(\topRetGroup,
\wtName^{[\dampFactor]})$ has edge weights defined as follows: for
every $\offDoc, \genDoc \in \topRetGroup$.
\[
\label{eq:smoothEdge}
\wtName^{[\dampFactor]}({\offDoc}\pointTo {\genDoc}) = (1-\dampFactor)\cdot \frac{1}{\vert \topRetGroup
  \vert} + \dampFactor \cdot 
\frac{\wtName({\offDoc}\pointTo{\genDoc})}{\sum_{\genDoc' \in \topRetGroup}    \wtName({\offDoc}\pointTo{\genDoc'})  },
\]
\end{definition}
The weights of
all edges leading out of any given node in  $\smoothG{G}{\dampFactor}$ sum to 1
and thus may be treated as {\em transition probabilities}.

With these concepts in hand, we can now phrase our
\prestigeText-determination task as follows: given a
\generationText graph, compute for each node (i.e., document)  how much
\prestigeText is ``transferred'' to it from other nodes ---  by our edge-weight definitions, centrality therefore corresponds to the degree to which
a document is responsible for ``\generating'' (perhaps indirectly) the
other documents in the initially retrieved set.  We now consider different ways to
formalize this notion of transferrence of \prestigeText.

\subsection{Computing Graph \PrestigeText}
\label{sec:induceprestige}

A straightforward way to define the \prestigeText of a document $\doc$
with respect to a given graph $G = (\topRetGroup,\wtName)$ is
to set it to $\doc$'s weighted in-degree, which we call its {\em  \wtedInDegree}:
\begin{equation}
\label{eq:wted-indegree}
\scoreFnG{\doc}{G} \definedas \sum_{\offDoc \in \topRetGroup} \edgeWeight{\offDoc}{\doc}.
\end{equation}
The \firstmention{\simpleCount} algorithm sets $G = \dgraphU$, so that
the only thing that matters is how many \offsText $\doc$ has; it is thus reminiscent of the
journal {\em impact factor} function  from bibliometrics
 \cite{Garfield:72a}, which computes normalized counts of explicit citation links.
The \firstmention{\weightedCount} algorithm sets $G = \dgraphW$, so
that the \generationText probabilities that $\doc$ assigns to its
\offText are factored in as well.

As previously noted by Pinski and Narin in their work on {\em
  influence weights}
\cite{Pinski+Narin:76a}, one intuition not accounted for by
weighted in-degree methods is that 
a document with 
even
a great many \offsText
should not be considered \prestigious (or relevant) if those \offsText are
themselves very non-\prestigious.  We can easily
modify Equation \ref{eq:wted-indegree} to model this
intuition; we simply 
{scale} the evidence from a particular \offText
document  by that \offText's \prestigeText ,
thus arriving at the following {\em recursive} equation:
\begin{equation}
\label{eq:basicPR}
\scoreFnGR{\doc}{G} \definedas \sum_{\offDoc \in \topRetGroup}
\edgeWeight{\offDoc}{\doc}\cdot \scoreFnGR{\offDoc}{G},
\end{equation}
where we also require that $\sum_{\doc \in \topRetGroup}
\scoreFnGR{\doc}{G} = 1$.
Unfortunately,  for
arbitrary $\dgraphU$ and $\dgraphW$,
Equation \ref{eq:basicPR}
may
not have a unique solution or even any solution at all
under the normalization constraint just given;
however,  a unique solution {\em is}
guaranteed to exist for their  PageRank-{smoothed}
versions.\footnote{The edge weights correspond to the transition
probabilities for a Markov
chain that is aperiodic and irreducible, and hence has a unique
stationary distribution \cite{Grimmett+Stirzaker:01a} that can be
computed by a variety of means
\cite{Stewart:94a,Golub+VanLoan:96a,Grassman+Taksar+Heyman:85a}.  
In our experiments, power iteration converged
very quickly.
}
By analogy with the two \wtedInDegree algorithms
given above, then, we have the \firstmention{\onlyUniformPrior} algorithm, which
sets $G = \smoothG{\dgraphU}{\dampFactor}$ and is a direct analog of PageRank, and the
\firstmention{\onlyPrior} algorithm, which sets $G=\smoothG{\dgraphW}{\dampFactor}$.

\subsection{Incorporating Initial Scores}
\label{sec:addLM}

The \prestigeText scores presented above can 
be used in
isolation as criteria by which to rank the documents in
$\topRetGroup$.
However, if available, it might 
be useful to
incorporate 
more information from the initial retrieval engine
to help handle cases where \prestigeText and relevance are not
strongly correlated.
(Recall that 
it participates in any case by specifying the
set $\topRetGroup$.)
In our experiments, we explore one concrete instantiation of this
approach: we apply language-model-based retrieval
\cite{Ponte+Croft:98a,Croft+Lafferty:03a} to determine $\topRetGroup$,
and consider the following family of re-ranking criteria:
\begin{equation}
\label{eq:basicLMRanking}
\scoreGeneric(\doc; G) \cdot \genprob{\doc}{\query},
\end{equation}
where  $\doc \in \topRetGroup$  and $\scoreGeneric$ is one of the \prestigeText functions defined in
the previous section.  
This gives rise to the algorithms \firstmention{\simpleCountPrior},
\firstmention{\weightedCountPrior}, \firstmention{\uniformRandomWalk},
and \firstmention{\weightedRandomWalk}. 

Incidentally, our 
choosing $\genprob{\doc}{\query}$ 
as initial score function
has the interesting
consequence that it suggests interpreting 
$\scoreGeneric(\doc; G)$ as
a document ``prior'' 
--- in fact, Lafferty and Zhai write, ``with
hypertext,  [a document prior]  might be the distribution calculated using
the `PageRank' scheme'' \cite{Lafferty+Zhai:01a}.  We will return to this idea later.

\subsection{Estimating \generationCaps Probabilities: Length and Entropy Effects}
\label{sec:lmspec}

\generationCaps probabilities form the basis for the graphs on which
our algorithms are defined.  This section describes our method for
estimating these probabilities.

Let
$\freq{\word}{\altArbGroupMember}$ denote the number of times the term $\word$
occurs in the text 
or text collection
$\altArbGroupMember$. 
What is often called 
the {\em maximum-likelihood
estimate} (MLE)
of $\word$
with respect to $\altArbGroupMember$
is defined as
$$\mlprobTerm{\altArbGroupMember}{\word} \definedas \frac{\freq{\word}{\altArbGroupMember}}{\sum_{\word'}
\freq{\word'}{\altArbGroupMember}}.$$
Some prior work in language-model-based retrieval
\cite{Liu+Croft:04a,Zhai+Lafferty:01a} employs
a {\em Dirichlet-smoothed} 
version:
\[
\dirichletLMTerm{\altArbGroupMember}{\word}{\dirichletParam} \definedas
\frac{\freq{\word}{\altArbGroupMember} +
\dirichletParam\cdot\mlprobTerm{\corpus}{\word}}{\sum_{\word'}
\freq{\word'}{\altArbGroupMember} + \dirichletParam};
\]
the smoothing parameter $\dirichletParam$ controls the degree of
reliance on relative frequencies in the corpus
rather than on the counts
in $\altArbGroupMember$. 
Both 
estimates just described
are typically extended to distributions over term sequences by
assuming that terms are independent: for an $\wordseqlength$-term
text sequence $\wordseq$, 
\begin{eqnarray*}
\mlprob{\altArbGroupMember}{\wordseq}
& \definedas &  \prod_{\wordIndex=1}^{\wordseqlength}
\mlprobTerm{\altArbGroupMember}{\word_\wordIndex}; \\
\dirichletLM{\altArbGroupMember}{\wordseq}{\dirichletParam} &
\definedas &
\prod_{\wordIndex=1}^{\wordseqlength}
\dirichletLMTerm{\altArbGroupMember}{\word_\wordIndex}{\dirichletParam}.
\end{eqnarray*}
Another estimation approach, which we adopt,  incorporates the Kullback-Leibler
divergence  $D$ between
document language models
\cite{Kurland+Lee:04a,Kurland+Lee+Domshlak:05a} (see also previously proposed ranking principles
\cite{Ng:00a,Lafferty+Zhai:01a}):
unless otherwise specified, 
for document $\doc$ and
word sequence $\wordseqvar$ (in our setting, either
a document or the query), we 
set $\genprob{\doc}{\wordseqvar}$ to
\begin{equation}
\genprobKL{\doc}{\wordseqvar} \definedas
\exp\left(-\kld{\mlprobTerm{\wordseqvar}{\cdot}}{\dirichletLMTerm{\doc}{\cdot}{\dirichletParam}}\right).
\label{eq:KLestimate}
\end{equation}

Equation \ref{eq:KLestimate} has some useful properties.
We can show that
$$\genprobKL{\doc}{\wordseqvar} =
\underbrace{(\dirichletLM{\doc}{\wordseqvar}{\dirichletParam})^{\frac{1}{\wordseqvarlength}}}_{\mbox{term
A}}
\cdot 
\underbrace{\exp(\entropy(\mlprobTerm{\wordseqvar}{\cdot}))}_{\mbox{term B}},
$$
where 
$\entropy$ is the entropy
function. 
Now,  observe that
for both $\bareMLProb{\altArbGroupMember}(\cdot)$ and $\bareDirProb{\altArbGroupMember}(\cdot)$, longer text
sequences tend to be assigned lower probabilities; this would correspond to
an unmotivated reduction of weights 
for edges out of long documents in 
the
graph  $\dgraphW$.  However, 
Term A length-normalizes $\dirichletLM{\doc}{\wordseqvar}{\dirichletParam}$ via the {\em geometric mean}, which has helped ameliorate
numerical problems in previous work \cite{Lavrenko+al:02a}.
Additionally, term B raises the \generation probability for 
texts with high-entropy MLE term distributions.
High entropy may be correlated with
a larger number of 
unique terms ---
for example, we get an entropy of 0 for the  document ``Salvador Salvador Salvador''
but $\log 3$ for ``Toronto Sheffield Salvador'' --- 
which, in turn,  has previously been suggested as a cue for relevance
\cite{Singhal+Buckley+Mitra:96a,Hiemstra+Kraaij:99a}.
Hence, \generatorsText of documents inducing high-entropy language models may be good
candidates for
\prestigeText status.
(We hasten to point out, though, that 
for the algorithms based on 
smoothed graphs (Definition \ref{defn:smoothGraph}), the entropy term
cancels out
due to 
our
normalization of edge weights.)

\section{Related Work}
\label{sec:relwork} 

Work on \rerankApproach in 
traditional ad hoc information retrieval has mainly focused on {\em
query-dependent clustering}, wherein one seeks to compute and exploit
a clustering of the initial retrieval results
\cite{Willett:85a,Hearst+Pedersen:96a,Leuski:01a,Tombros+Villa+Rijsbergen:02a,Liu+Croft:04a}.
Clusters represent structure within a document set, but do not
directly induce an obvious single criterion or principle by which to
rank documents; for instance, they have been  used to improve
rankings indirectly by serving as smoothing
mechanisms \cite{Liu+Croft:04a}.  Interestingly, some \prestigeText measures have been
previously employed to produce clusterings \cite{Tishby+Slonim:00a}.

There has been
increasing use of techniques based on graphs induced
by implicit relationships between 
documents or other linguistic items
\cite{Hatzivassiloglou+McKeown:97a,Dhillon:01a,Joachims:03a,Erkan+Radev:04b,Mihalcea+Tarau:04a,Pang+Lee:04a,Toutanova+Manning+Ng:04a}.
The work in the domain of text summarization
\cite{Erkan+Radev:04b,Mihalcea+Tarau:04a} most resembles ours,
in that 
it also computes \prestigeText on graphs 
 (although the nodes correspond to sentences 
or terms
instead
of documents).  
Perhaps the main contrast with our work is that links
were not induced by \generationText probabilities; 
Section
\ref{sec:exp} presents the results of experiments studying the
relative merits of our particular choice of link definition.

Our \prestigeText scores constitute a relationship-based re-ranking
criterion that can serve as a bias affecting the initial retrieval
engine's scores, as in Equation \ref{eq:basicLMRanking}.  Alternative
biases that are based on individual documents alone have also been
investigated.  Functions 
incorporating document or average word length
\cite{Hiemstra+Kraaij:99a,Kraaij+Westerveld:00a,Miller+Leek+Schwarz:99a}
are applicable 
in our setting; we report on experiments with (variants
of) document length in 
Section \ref{sec:exp}.  Other previously
suggested biases that may be somewhat less appropriate for general
domains include document source \cite{Miller+Leek+Schwarz:99a} and
creation time
\cite{Li+Croft:03a},  and webpage hyperlink in-degree and URL form
\cite{Kraaij+Westerveld+Hiemstra:02a}.

\section{Evaluation}
\newcommand{\better}[1]{\mathit{#1}}

\label{sec:expSet}

\subsection{Experimental Setting}

The objective of \rerankApproach is to (re-)order an
initially-retrieved document set $\topRetGroup$ so as to improve precision at the very
top ranks of the final results.  
Therefore, we employed the following three evaluation metrics: the
precision of the top 5 documents (prec@5), the precision of the top 10
documents (prec@10), and the mean reciprocal rank of the first relevant document
(MRR) \cite{Shah+Croft:04a}.

We are interested in the general validity of the various
\rerankApproach methods we have proposed.  We believe that a good way
to 
emphasize the effectiveness (or lack thereof) of the underlying principles is to
downplay the role of parameter tuning.
Therefore, we made the
following design decisions, with the effect that {\em the performance
numbers we report 
are purposely not necessarily the 
 best achievable by
exhaustive parameter search}:
\begin{itemize}
\item The {\em initial ranking} that created the set $\topRetGroup$
  was built according to the function 
$\genprobKL{\doc}{\query}$ where
  the value of $\dirichletParam$ was chosen to optimize the non-interpolated average precision of the top 1000 retrieved
documents.  This is {\em not} one of our evaluation metrics, but is a
  reasonable general-purpose optimization criterion. (In fact, results
  with this initial ranking turned out to be statistically indistinguishable from
  the results obtained by optimizing with respect to the actual
  evaluation metrics, although of course they were lower in absolute terms.)
\item 
We {only} optimized settings 
for $\numGenerators$
(the ancestry parameter controlling the 
number of top \generatorsText considered for each document) and 
$\dampFactor$ (the edge-weight smoothing factor)
with respect to  precision among the top 5 documents, not 
with respect
to all three evaluation
metrics employed.
\end{itemize}

The search ranges for the latter two parameters were:
\begin{center}
\begin{tabular}{ll}
$\numGenerators$: & $4,9,19,\ldots,\numretdocs-1$ \\
$\dampFactor$: & $0,0.05,0.1,0.2,\ldots,0.9,0.95$
\end{tabular}
\end{center}
As it turned out, for many instances 
(except for the \weightedCount algorithm), the optimal value of 
$\numGenerators$
with respect to precision at 5 
was
either $4$ or $9$, suggesting that a relatively small number of
\generatorsText 
per document 
should be considered
when constructing the graph. 
In contrast, 
$\dampFactor$ 
exhibited substantial variance in 
optimal 
value for precision at 5 
in some of our datasets.
We set $|\topRetGroup|$, the number of
initially-retrieved documents, to $50$ in all results reported below
(similar 
performance patterns were
obtained when  $|\topRetGroup| = 100$).

The remaining details are as follows.
We conducted our experiments on the following four TREC corpora:
\begin{center}
\begin{tabular}{|lrlc|}\hline
corpus & \# of docs & \multicolumn{1}{c}{queries} & disk(s)  \\ \hline
AP89  & 84,678 & 1-46,48-50 & 1 \\ 
AP & 242,918 &  51-64, 66-150 & 1-3 \\
WSJ & 173,252 & 151-200 & 1-2  \\ 
TREC8 & 528,155 & 401-450 & 4-5\\ \hline
\end{tabular}
\end{center}
(AP89 is a subset of AP containing articles just from the year 1989).
All documents and queries (in our case, TREC-topic titles) were
stemmed using the Porter stemmer
and tokenized, but
no other 
pre-processing 
steps were applied.
We used the Lemur toolkit \cite{Ogilvie+Callan:2001}
for language-model estimation.  
Statistically-significant differences in performance 
were determined 
using
the two-sided Wilcoxon test at a confidence level of $95\%$.

\subsection{Results}
\label{sec:exp}

\newcommand{\baseTableSize}{\footnotesize}
\newcommand{\baseTableCaption}{
Primary experimental results, showing algorithm performance with respect
to our 12 evaluation settings (3 performance metrics $\times$ 4
corpora). For each evaluation
setting,  improvements over the \underline{optimized} baselines are given in {italics};
statistically significant differences between our
\rerankApproach algorithms and the initial ranking and optimized baselines are indicated by {\em i} and {\em o} respectively;
{\bf bold}  highlights
the best results over all ten algorithms.

\hspace*{.1in} Notice that even though the \rerankApproach algorithms were optimized
for prec@5 only (and produce the best results for this metric), they
still perform well with respect to the other two metrics.
}
\begin{table*}[t]
\begin{flushleft}
\hspace*{-0.25in}
\baseTableSize
\begin{tabular}{|l||c|c|c|c|c|c|c|c|c|c|c|c|}
\hline
& \multicolumn{3}{|c|}{AP89}& \multicolumn{3}{|c|}{AP}& \multicolumn{3}{|c|}{WSJ}& \multicolumn{3}{|c|}{TREC8}\\ \cline{2-13} 
 & {prec$@5$} & {prec$@10$} & {MRR} & {prec$@5$} & {prec$@10$} & {MRR} & {prec$@5$} & {prec$@10$} & {MRR} & {prec$@5$} & {prec$@10$} & {MRR} \\ \hline
upper bound & $ 63.7 $ $ ^{}_{}$& $ 53.1 $ $^{}_{}$& $ 75.5 $ $^{}_{}$& $ 87.6 $ $ ^{}_{}$& $ 78.8 $ $^{}_{}$& $ 93.0 $ $^{}_{}$& $ 89.6 $ $ ^{}_{}$& $ 80.0 $ $^{}_{}$& $ 100.0 $ $^{}_{}$& $ 94.4 $ $ ^{}_{}$& $ 85.0 $ $^{}_{}$& $ 98.0 $ $^{}_{}$\\ \hline
\abbrevInit & $ 28.3 $ $ ^{}_{}$& $ 26.5 $ $^{}_{}$& $ 52.3 $ $^{}_{}$& $ 45.7 $ $ ^{}_{}$& $ 43.2 $ $^{}_{}$& $ 59.6 $ $^{}_{}$& $ 54.8 $ $ ^{}_{}$& $ 48.4 $ $^{}_{}$& $ 76.2 $ $^{}_{}$& $ 50.0 $ $ ^{}_{}$& $ 45.6 $ $^{}_{}$& $ 69.1 $ $^{}_{}$\\ \hline
opt. baselines & $ 30.0 $ $ ^{}_{}$& $ 27.4 $ $^{}_{}$& \mbox{\boldmath$ 54.3 $} $^{}_{}$& $ 46.5 $ $ ^{}_{}$& $ 43.9 $ $^{}_{}$& $ 63.5 $ $^{}_{}$& $ 56.0 $ $ ^{}_{}$& $ 49.4 $ $^{}_{}$& $ 77.2 $ $^{}_{}$& $ 51.2 $ $ ^{}_{}$& $ 46.4 $ $^{}_{}$& \mbox{\boldmath$ 69.6 $} $^{}_{}$\\ \hline
\hline
\hline
\abbrevVote & $ 29.6 $ $ ^{}_{}$& $ \better{27.8}$ $^{}_{}$& $ 39.5 $ $^{}_{o}$& $ \better{50.9}$ $^{}_{}$& $ \better{49.0}$ $^{i}_{o}$& \mbox{\boldmath$ 66.3 $} $^{}_{}$& $ 50.0 $ $ ^{}_{}$& $ 46.6 $ $^{}_{}$& $ 66.7 $ $^{}_{}$& $ 50.0 $ $ ^{}_{}$& $ 45.0 $ $^{}_{}$& $ 62.0 $ $^{}_{}$\\ \hline
\abbrevWeightedVote & $ \better{31.3}$ $^{}_{}$& $ \better{29.6}$ $^{}_{}$& $ 46.8 $ $^{}_{}$& $ \better{51.3}$ $^{}_{}$& $ \better{48.7}$ $^{i}_{}$& $ \better{64.4}$ $^{}_{}$& $ 52.0 $ $ ^{}_{}$& $ 47.8 $ $^{}_{}$& $ 63.3 $ $^{}_{o}$& $ 49.2 $ $ ^{}_{}$& $ 43.4 $ $^{}_{}$& $ 63.7 $ $^{}_{}$\\ \hline
\abbrevSimpleVotePrior & \mbox{\boldmath$ 33.5 $} $^{}_{}$& $ 27.0 $ $^{}_{}$& $ 46.5 $ $^{}_{}$& $ \better{51.3}$ $^{i}_{}$& \mbox{\boldmath$ 49.4 $} $^{i}_{o}$& $ 63.2 $ $^{}_{}$& $ \better{56.4}$ $^{}_{}$& $ 49.2 $ $^{}_{}$& $ 73.6 $ $^{}_{}$& $ \better{52.8}$ $^{}_{}$& \mbox{\boldmath$ 52.0 $} $^{i}_{o}$& $ 66.6 $ $^{}_{}$\\ \hline
\abbrevWeightedVotePrior & $ \better{31.7}$ $^{}_{}$& $ \better{27.6}$ $^{}_{}$& $ 48.4 $ $^{}_{}$& $ \better{51.1}$ $^{i}_{}$& $ \better{48.4}$ $^{i}_{o}$& $ 63.0 $ $^{}_{}$& $ \better{57.2}$ $^{}_{}$& $ \better{50.0}$ $^{}_{}$& $ 77.2 $ $^{}_{}$& $ \better{51.6}$ $^{}_{}$& $ \better{49.6}$ $^{i}_{}$& $ 64.5 $ $^{}_{}$\\ \hline
\hline
\hline
\abbrevOnlyUniformPrior & $ \better{31.3}$ $^{}_{}$& $ \better{28.9}$ $^{}_{}$& $ 46.4 $ $^{}_{}$& $ \better{51.5}$ $^{}_{}$& $ \better{48.9}$ $^{i}_{}$& $ 63.4 $ $^{}_{}$& $ 53.6 $ $ ^{}_{}$& $ \better{49.6}$ $^{}_{}$& $ 68.5 $ $^{}_{}$& $ \better{52.0}$ $^{}_{}$& $ 44.6 $ $^{}_{}$& $ 66.5 $ $^{}_{}$\\ \hline
\abbrevOnlyPrior & $ \better{32.2}$ $^{}_{}$& $ \better{29.6}$ $^{}_{}$& $ 40.5 $ $^{}_{o}$& $ \better{52.1}$ $^{i}_{}$& $ \better{49.1}$ $^{i}_{o}$& $ \better{63.9}$ $^{}_{}$& $ 54.0 $ $ ^{}_{}$& $ 49.2 $ $^{}_{}$& $ 70.2 $ $^{}_{}$& $ \better{52.4}$ $^{}_{}$& $ 44.6 $ $^{}_{}$& $ 66.5 $ $^{}_{}$\\ \hline
\abbrevUniformRandomWalk & $ \better{33.0}$ $^{}_{}$& $ \better{29.3}$ $^{}_{}$& $ 45.8 $ $^{}_{}$& $ \better{52.1}$ $^{i}_{o}$& $ \better{49.2}$ $^{i}_{o}$& $ \better{64.3}$ $^{}_{}$& \mbox{\boldmath$ 58.8 $} $^{i}_{}$& \mbox{\boldmath$ 51.0 $} $^{i}_{}$& \mbox{\boldmath$ 78.6 $} $^{}_{}$& $ \better{55.6}$ $^{}_{}$& $ 46.0 $ $^{}_{}$& $ 68.4 $ $^{}_{}$\\ \hline
\abbrevWeightedRandomWalk & \mbox{\boldmath$ 33.5 $} $^{}_{}$& \mbox{\boldmath$ 29.8 $} $^{}_{}$& $ 46.0 $ $^{}_{}$& \mbox{\boldmath$ 52.9 $} $^{i}_{o}$& $ \better{49.0}$ $^{i}_{o}$& $ 62.6 $ $^{}_{}$& \mbox{\boldmath$ 58.8 $} $^{i}_{}$& $ \better{50.6}$ $^{}_{}$& \mbox{\boldmath$ 78.6 $} $^{}_{}$& \mbox{\boldmath$ 56.0 $} $^{}_{}$& $ 45.8 $ $^{}_{}$& $ 67.6 $ $^{}_{}$\\ \hline
\end{tabular}
\caption{\label{tab:algTable,baseMetric=kl,prMetric=lm,optMeasure=prec5} \baseTableCaption}
\end{flushleft}
\end{table*}

In the tables that follow, we use the following abbreviations for
algorithm names.
\begin{center}
\begin{tabular}{|ll|} \hline
\abbrevVote & \simpleCount \\
\abbrevWeightedVote & \weightedCount\\ \hline
\abbrevOnlyUniformPrior & \onlyUniformPrior\\
\abbrevOnlyPrior & \onlyPrior\\ \hline \hline
\abbrevSimpleVotePrior & \simpleCountPrior\\
\abbrevWeightedVotePrior & \weightedCountPrior\\ \hline
\abbrevUniformRandomWalk & \uniformRandomWalk\\
\abbrevWeightedRandomWalk & \weightedRandomWalk \\\hline
\end{tabular}
\end{center}

\subsubsection{Primary evaluations}

Our main experimental results are presented in Table
\ref{tab:algTable,baseMetric=kl,prMetric=lm,optMeasure=prec5}. 
The first three rows specify reference-comparison data.
The initial ranking was, as described above, produced using
$\genprobKL{\doc}{\query}$ with $\dirichletParam$ chosen to optimize for
non-interpolated precision at 1000.  
The {\em empirical upper bound on \rerankApproach},
which applies to any algorithm that re-ranks $\topRetGroup$, 
indicates the performance that would be achieved if all the relevant
documents within the initial fifty  were placed at the top of the
retrieval list:
note that these bounds indicate that the initial rankings for AP89 are
quite worse than those for the other three corpora.
We also computed an
{\em optimized baseline} for each  metric $m$ and test corpus $\corpus$;
this consists of ranking all the documents (not just those in $\topRetGroup$) by
$\genprobKL{\doc}{\query}$, with
$\dirichletParam$ chosen to yield the best $m$-results on $\corpus$.
As a sanity check, we observe that the performance of the initial
retrieval method is always below that of the corresponding optimized
baseline (though not statistically distinguishable from it).

The first question we are interested in is how our \rerankApproach
algorithms taken as a whole do.
As shown in Table \ref{tab:algTable,baseMetric=kl,prMetric=lm,optMeasure=prec5},
our methods  
improve upon the initial ranking 
in many cases,
specifically, roughly 2/3
of the 96 relevant comparisons (8 \prestigeText-based algorithms
$\times$ 4 corpora $\times$ 3 evaluation metrics).
An even more gratifying observation is that Table
\ref{tab:algTable,baseMetric=kl,prMetric=lm,optMeasure=prec5} shows
(via italics and boldface) that in many cases, 
our algorithms, even though optimized for precision at 5, can
outperform a language model optimized for a different (albeit related)
metric $m$ even when performance is measured with respect to $m$;
see, for example, the results for precision at 10 
on the AP corpus.

Closer examination of the
results in Table
\ref{tab:algTable,baseMetric=kl,prMetric=lm,optMeasure=prec5} reveals
that %
in 
about 60\% of the
48 relevant comparisons,
our algorithms not only are at least as effective when applied
to the graph $\dgraphW$ 
as  when applied 
to $\dgraphU$, but often yield better performance results;
the comparison between
\onlyPrior (\abbrevOnlyPrior) and 
\onlyUniformPrior (\abbrevOnlyUniformPrior) is a  good
example.  
These results imply that it is a bit better to
explicitly incorporate \generationText probabilities into the edge
weights of our \generationText graphs  than to treat all the top
\generatorsText of a document equally.

Another observation 
we can draw from Table
\ref{tab:algTable,baseMetric=kl,prMetric=lm,optMeasure=prec5}
is that
adding in
query-\generationText probabilities as
weights on the \prestigeText scores 
(see Equation \ref{eq:basicLMRanking})
tends to 
enhance performance. This can be seen by comparing  rows labeled with some
algorithm abbreviation ``{X}'' against the corresponding rows labeled
``\abbrevAsPrior{{X}}'': 
about 80\% of the 48 relevant comparisons exhibit this improvement.
Most of the counterexamples occur in settings involving 
precision at 10 and MRR, which we did not optimize our algorithms for.

Similarly, by comparing ``Y''-labeled rows with ``\scaledAbbrev-Y''-labeled ones, we
see that 
in about 70\% of the 48 relevant comparisons, 
it 
is  better 
to use the recursive
formulation of Equation \ref{eq:basicPR}, where  
the \prestigeText of a document is affected by the \prestigeText of its
\offText, 
than to ignore 
\offText \prestigeText as is
done by Equation \ref{eq:wted-indegree}.

Perhaps not surprisingly, then, the \uniformRandomWalk and
\weightedRandomWalk algorithms, which combine the
two preferred features just described 
(recursive \prestigeText computation and
use of the initial search engine's score function)
appear
to be our best performing
algorithms: working from a starting point below the optimized
baselines, they improve the initial retrieval set to yield results
that even at their worst, are 
not only clearly better than the initial ranking for precision
at 5 and 10, but are also
merely
statistically indistinguishable from the optimized baselines.
Moreover, 
in one setting  (AP, precision at 10) they actually produce
statistically significant improvements over the optimized baseline
even though they were not optimized for that
evaluation metric.

It is interesting to note that the {\em relative} performance of our
algorithms does not seem to depend strongly on the quality of the
initial ranking, in the following sense.
The average percentage of relevant documents among the 50
that are initially retrieved is $21\%$,
$35.5\%$, $33.3\%$ and $30.3\%$ for AP89, AP, WSJ and TREC8,
respectively, but the
relative improvements for precision at 5 and 10 that our algorithms
achieve with  respect to the initial ranking are almost always higher on AP89
than on WSJ or TREC8.

\newcommand{\cosTableSize}{}
\newcommand{\cosTableCaption}[1]{\RerankApproach based on language models
(LM) vs. \rerankApproach based on cosine-measured vector-space
proximity (VEC).  We indicate the settings in which the relative
difference was at least $#1\%$
with
either a ``\A''  (LM superior) or a
``\F'' (VEC superior).}
{
\begin{table*}[t]
\begin{center}
\cosTableSize\begin{tabular}{ll|*8{c}} \\
& & \abbrevVote & \abbrevWeightedVote & \abbrevSimpleVotePrior & \abbrevWeightedVotePrior & \abbrevOnlyUniformPrior & \abbrevOnlyPrior & \abbrevUniformRandomWalk & \abbrevWeightedRandomWalk \\ \hline \hline
& prec @5& \F & \Dm & \C & \C & \C & \Dm & \C & \C \\ {\bf AP89} 
& prec @10& \C & \Am & \C & \C & \C & \C & \C & \Am \\
& MRR& \X & \F & \F & \C & \F & \X & \F & \C \\ \hline& prec @5& \Am & \Am & \Am & \Am & \Am & \Am & \A & \A \\ {\bf AP} 
& prec @10& \A & \Am & \A & \A & \Am & \A & \A & \A \\
& MRR& \Am & \Am & \C & \Am & \C & \Am & \Am & \C \\ \hline& prec @5& \C & \C & \C & \C & \C & \Amm & \C & \C \\ {\bf WSJ} 
& prec @10& \C & \C & \C & \C & \Am & \Amm & \C & \C \\
& MRR& \C & \X & \C & \C & \C & \C & \C & \C \\ \hline& prec @5& \Amm & \Amm & \C & \C & \Am & \Am & \Am & \Am \\ {\bf TREC8} 
& prec @10& \Amm & \Amm & \B & \Am & \C & \Amm & \C & \C \\
& MRR& \F & \C & \C & \F & \C & \C & \C & \C \\ \hline\end{tabular}
\caption{\label{tab:algTable,compareKLandCosine,baseMetric=kl,optMeasure=prec5,autoGen} \cosTableCaption{5}}
\end{center}
\end{table*}
}

\newcommand{\priorTableSize}{\footnotesize}
\newcommand{\priorTableCaption}{Comparison 
between our use of language-model-based structural-\prestigeText
scores in Equation \protect\ref{eq:basicLMRanking} vs. non-structural
re-ranking heuristics.
For each evaluation
setting,  {italics} mark improvements over the default baseline of uniform
\prestigeText scores, 
stars (*) indicate statistically significant differences with this default baseline, 
and {\bf bold}  highlights
the best results over all eight algorithms.}
\begin{table*}[t]
\begin{flushleft}
\hspace*{-0.25in}
\priorTableSize
\begin{tabular}{|l||c|c|c|c|c|c|c|c|c|c|c|c|}
\hline
& \multicolumn{3}{|c|}{AP89}& \multicolumn{3}{|c|}{AP}& \multicolumn{3}{|c|}{WSJ}& \multicolumn{3}{|c|}{TREC8}\\ \cline{2-13}
& {prec$@5$} & {prec$@10$} & {MRR} & {prec$@5$} & {prec$@10$} & {MRR} & {prec$@5$} & {prec$@10$} & {MRR} & {prec$@5$} & {prec$@10$} & {MRR} \\ \hline
\abbrevUniformPrior & $ 28.3  $& $ 26.5  $& $ 52.3  $& $ 45.7  $& $ 43.2  $& $ 59.6  $& $ 54.8  $& $ 48.4  $& $ 76.2  $& $ 50.0  $& $ 45.6  $& $ 69.1  $\\ \hline
\hline
\abbrevWeightedVote & $ \better{31.7 }  $& $ \better{27.6 }  $& $ 48.4  $& $ \better{51.1 *}  $& $ \better{48.4 *}  $& \mbox{\boldmath$ 63.0  $}& $ \better{57.2 }  $& $ \better{50.0 }  $& $ \better{77.2 }  $& $ \better{51.6 }  $& \mbox{\boldmath$ 49.6 * $}& $ 64.5  $\\ \hline
\abbrevOnlyPrior & \mbox{\boldmath$ 33.5  $}& \mbox{\boldmath$ 29.8  $}& $ 46.0  $& \mbox{\boldmath$ 52.9 * $}& \mbox{\boldmath$ 49.0 * $}& $ \better{62.6 }  $& \mbox{\boldmath$ 58.8 * $}& \mbox{\boldmath$ 50.6  $}& \mbox{\boldmath$ 78.6  $}& \mbox{\boldmath$ 56.0  $}& $ \better{45.8 }  $& $ 67.6  $\\ \hline
\hline
length & $ \better{29.1 }  $& $ 24.3  $& $ 50.8  $& $ 41.6  $& $ 41.4  $& $ 55.3  $& $ 44.4 * $& $ 42.4 * $& $ 64.6 * $& $ 47.2  $& $ 41.4  $& $ 64.2  $\\ \hline
log(length) & $ \better{30.4 }  $& $ \better{27.0 }  $& $ \better{52.5 }  $& $ 45.3  $& $ 43.2  $& $ \better{60.6 }  $& $ \better{57.2 }  $& $ \better{49.0 }  $& $ 69.8 * $& $ 49.6  $& $ \better{46.8 }  $& $ \better{69.2 }  $\\ \hline
entropy & $ \better{30.0 }  $& $ 26.5  $& \mbox{\boldmath$ 52.6  $}& $ \better{46.1 }  $& $ 42.5  $& $ \better{60.8 }  $& $ \better{56.8 }  $& $ \better{48.6 }  $& $ 71.1 * $& $ 49.6  $& $ \better{46.8 }  $& \mbox{\boldmath$ 71.7 * $}\\ \hline
uniqTerms & $ 27.4  $& $ 24.8  $& $ 52.3  $& $ 42.0  $& $ 41.3  $& $ 56.2  $& $ 50.0  $& $ 44.6  $& $ 68.8  $& $ 49.2  $& $ 44.2  $& $ \better{71.2 }  $\\ \hline
log(uniqTerms) & $ \better{30.4 }  $& $ \better{27.0 }  $& $ \better{52.5 }  $& $ \better{45.9 }  $& $ 42.3  $& $ \better{60.8 }  $& $ \better{57.2 }  $& $ \better{49.0 }  $& $ 70.0 * $& $ 49.6  $& $ \better{47.2 }  $& $ \better{70.0 }  $\\ \hline
\end{tabular}
\caption{\label{tab:priorTable,baseMetric=kl,prMetric=lm,optMeasure=prec5} \priorTableCaption}
\end{flushleft}
\end{table*}

\subsubsection{Links based on the vector-space model}
\label{sec:cos}
We have advocated the use of \generationText relationships to define
\prestigeText, where these 
asymmetric relationships are
based on language-model probabilities.  However, other inter-document
relationships have been previously exploited in information retrieval.
Perhaps the most well-known is {\em vector-space proximity}, with the cosine frequently used as (symmetric) closeness metric;
indeed, as mentioned above, previous work in summarization
\cite{Erkan+Radev:04b} has used the cosine to
determine \prestigeText in ways very similar to the ones we have
considered.  It is thus important to examine whether the performance
improvements we have achieved can be reproduced, or even surpassed, by
the use of vector-space-based links rather than language-model-based
\generationText links.

To run this evaluation, we simply modified Definition \ref{defn:docEnv} 
and all eight of our
\rerankApproach algorithms
to use the cosine of the angle between log tf.idf document vectors, rather than
language-model probabilities, to 
form the basis for determining the
edge weights of our
graphs.  
(Note that the fact that the cosine is symmetric does not imply that
edges $(\nodeFrom,\nodeTo)$ and  $(\nodeTo,\nodeFrom)$ get the same
weight even in our non-smoothed graphs --- document $\doc_1$ being a
top ``generator'' of $\doc_2$  with respect to the cosine does not
imply the reverse.)
It should be observed that the 
language-model weights on \prestigeText scores  (i.e., the
$\genprob{\doc}{\query}$
term in Equation
\ref{eq:basicLMRanking}, on which the 
``\asPrior{}''
algorithms are based) were {\em
not} replaced with cosine values, which makes sense since we want our comparison
 to  focus on the effect of different means of computing graph-based \prestigeText.

Table
\ref{tab:algTable,compareKLandCosine,baseMetric=kl,optMeasure=prec5,autoGen}
depicts the relative performance differences between using our
language-model-based graphs
and graphs induced using 
vector-space proximity
in the
manner just described.  For each choice of
algorithm, evaluation measure, and dataset, we indicate which
formulation, if any,
resulted in at least $5\%$ relative improvement with respect to the
other.
As can be seen, in at least three of our four corpora,
our language-modeling approach seems to be a more effective basis for
determining document \prestigeText
than 
the vector-space/cosine.
We hasten to point out, though, that in most
instances,
vector-space proximity yielded
better performance than the corresponding baselines
(the results are omitted 
since the precise numerical comparison does not yield additional information); this finding
provides further support to the idea that the overall \rerankApproach
approach is a  flexible and effective paradigm 
that can incorporate
different types of inter-document relationships when appropriate.

\subsubsection{Inducing \prestigeText with the HITS algorithm}

One well-known alternative method for computing \prestigeText in a
graph is the HITS algorithm \cite{Kleinberg:99a}, originally proposed
for Web search.  
There has been some work utilizing it for text
summarization in non-Web domains as well \cite{Mihalcea:04a}.  The
reason we have not yet discussed it in detail is that it differs
conceptually from our proposed algorithms in an important way:
two different 
notions of \prestigeText are identified, represented by {\em hub} and
{\em authority} scores.
While the concepts of hubs and authorities are highly suitable for
Web-search scenarios, it is less clear whether it is useful in our
setting to distinguish between the two.

As a preliminary investigation, 
we experimented with using hub and
authority scores as measures of \prestigeText on the \generationText
graphs we built.  Space constraints preclude a detailed discussion,
but the results may be summarized as follows.  We found that authority
scores yielded better performance than hub scores, and that the
results were generally at least as good as or better than those for
the optimized baselines.  However, they were slightly inferior in
several cases to those of the corresponding \wtedInDegree algorithms.
Thus, it seems that our method for graph construction can support a
variety of different algorithms, but that the HITS-style
hubs/authorities distinction may not be effective for the task we have
addressed.

\subsubsection{Non-\rerankApproach}

So far, we have discussed the use of graph-based \prestigeText as a
re-ranking criterion, the idea being that relationships between
documents can serve as an additional source of information.
Our best empirical results seem
to be produced by using the 
weighted 
formulation 
given in Equation
\ref{eq:basicLMRanking} from Section \ref{sec:addLM}:
\[
 \scoreGeneric(\doc; \arbGraph) \cdot 
\genprob{\doc}{\query}.
\]
Since, as noted above, in this equation
$\scoreGeneric(\doc; \arbGraph)$ can be regarded as a ``prior'' on
documents, it is natural to ask whether other
previously-proposed biases on \generationText probabilities might
prove similarly useful. The comparison is especially interesting because
these biases have tended to be 
isolated-document
heuristics; we thus refer to their use as a replacement for
$\scoreGeneric(\doc; \arbGraph)$ 
as ``non-\rerankApproach''.

Document length has been employed several times in the past
to model the intuition that longer texts contain more information
\cite{Hiemstra+Kraaij:99a,Kraaij+Westerveld:00a,Miller+Leek+Schwarz:99a}.
We refine this hypothesis to disentangle several distinct notions of
information: the number of {\em tokens} in a document, the {\em
distribution} of these {tokens}, and the number of 
{\em types} (``Salvador Salvador Salvador'' contains
three tokens but only one type).
Thus, as substitutions for
\prestigeText in the above expression, we consider not only document
length, but also the
{entropy} of the term distribution and the number of unique terms
(used as the basis for pivoted unique normalization in
\cite{Singhal+Buckley+Mitra:96a}).  As baseline, we took the initial
retrieval results; note that 
doing so corresponds to using a uniform
bias, or, equivalently, using no bias at all.

As can be seen in Table
\ref{tab:priorTable,baseMetric=kl,prMetric=lm,optMeasure=prec5}, 
taking the log of token or type count is an improvement over using the
raw frequencies, often yielding above-baseline performance.  The
entropy is more effective than raw frequency of either tokens or
types, and in two cases leads to the best performance overall.
However, in the majority of settings, \rerankApproach gives the
highest accuracies.

\subsubsection{Re-ranking vs. ranking}

We posed our \prestigeText-computation techniques as methods for
improving the results returned by an initial retrieval engine, and
showed that they are successful at accomplishing this goal.  But one
can ask whether it is necessary to restrict our attention to an
initial pool  $\topRetGroup$; that is, would we expect similarly good results if we
based our \generationText graphs on the entire corpus?
As it happens, preliminary experiments
with the \uniformRandomWalk and
\weightedRandomWalk algorithms 
on two {\em full} corpora (AP89 and LA combined with FR)
showed 
that one would be better off sticking with
the standard language-modeling approach if no pre-filtering of
documents is available.

We do not see this finding as surprising, for 
our intuition is that in the re-ranking case, there is a more
direct connection between  \prestigeText and relevance since we can
assume that relevant documents comprise a reasonable fraction of the
initial retrieval results.

\section{Conclusion}

We have proposed and evaluated a number of methods for \rerankApproach
using inter-document \generationText relationships based on language models.  Our main
experiments showed that even {\em non-optimized} instantiations of our
overall approach yield results rivaling 
those of {\em optimized}
baselines.  Further analysis revealed that  \generationText
relationships seem  more effective within our
\prestigeText-computation framework than relationships based on
vector-space proximity do, and that using inter-document relationships
seems to be a promising alternative to 
employing the isolated-document
heuristics we implemented
(several of which were novel to this study).
Based on our results, we believe that exploring other methods for
combining statistical language models and explicitly graph-based
techniques is a fruitful line for future research.

\paragraph*{Acknowledgments}
We thank James Allan, Bruce Croft, Carmel Domshlak, Jon
Kleinberg, Fernando Pereira and the anonymous reviewers for valuable discussions and comments. We also thank CMU for its hospitality during the year.
This paper is based upon work supported in part by the National
Science Foundation under grant no.  IIS-0329064 and CCR-0122581;
SRI International under subcontract no. 03-000211 on their project
funded by the Department of the Interior's National Business Center;
and 
 an Alfred P. Sloan Research Fellowship. Any opinions, findings,
and conclusions or recommendations expressed are those of the authors
and do not necessarily reflect the views or official policies, either
expressed or implied, of any sponsoring institutions, the
U.S. government, or any other entity.

\end{document}